\documentclass
[nofootinbib,superscriptaddress,notitlepage,twocolumn,pre]{revtex4-1}%
\usepackage{amssymb}
\usepackage{amsfonts}
\usepackage{amsmath}
\usepackage{graphicx}
\usepackage{color}
\usepackage{version}
\usepackage{numinsec}%
\providecommand{\U}[1]{\protect \rule{.1in}{.1in}}

\includeversion{Commentskip}
\includeversion{Commentdel}
\includeversion{Commentnew}
\newcommand{\BSKIP}{\  \color{red}$\Downarrow$SKIP$\Rightarrow$\ }
\newcommand{\ESKIP}{\  \color{red}$\Leftarrow$SKIP$\Uparrow$\color{black}\ }
\newcommand{\BDEL}{\  \color{cyan}$\Downarrow$DEL$\Rightarrow$\ }
\newcommand{\EDEL}{\  \color{cyan}$\Leftarrow$DEL$\Uparrow$\color{black}\ }
\newcommand{\BNEW}{\  \color{blue}$\Downarrow$NEW$\Rightarrow$\ }
\newcommand{\ENEW}{\  \color{blue}$\Leftarrow$NEW$\Uparrow$\color{black}\ }
\begin{document}
\title{Variational Truncated Wigner Approximation }
\author{Dries Sels}
\email{dries.sels@uantwerpen.be}
\author{Fons Brosens}
\email{fons.brosens@uantwerpen.be}
\affiliation{Physics Department, University of Antwerp, Universiteitsplein 1, 2060
Antwerpen, Belgium}
\date{\today}

\begin{abstract}
In this paper we reconsider the notion of an optimal effective Hamiltonian for
the semiclassical propagation of the Wigner distribution in phase space. An
explicit expression for the optimal effective Hamiltonian is obtained in the
short time limit by minimizing the Hilbert-Schmidt distance between the
semiclassical approximation and the real state of the system. The method is
illustrated for the quartic oscillator.

\end{abstract}
\maketitle

\section{Introduction\label{sec:Intro}}

Since the seminal work by Wigner~\cite{cit:Wigner}, phase space methods have
attracted considerable theoretical attention. The interest in phase space
methods has only increased over the years and it has found important
applications in the fields of quantum optics~\cite{cit:Glauber, cit:Sudershan,
cit:GardnerZoller}, atomic physics~\cite{cit:Harochecat}, cold
atoms~\cite{cit:Steel} or other bosonic systems~\cite{cit:Polkovnikov2010,
cit:Michiel}. Significant theoretical attention has been attracted by the
semiclassical (truncated Wigner) approximation of the full quantum phase space
dynamics. The full quantum dynamics can be represented in terms of quantum
jumps around the classical (symplectic) phase space flow. Within the
semiclassical truncated Wigner approximation one neglects the quantum jumps.
Consequently the quantum dynamics is approximated by the classical evolution
of the system. However, the initial state still caries the signatures of
statistics into the time-dependent solution. The justification for this
expansion around the classical trajectory is based on the observation that
every quantum jump carries an additional factor $\hbar^{2}~$%
\cite{cit:Polkovnikov2010}, which is obvious from the Moyal
expansion~\cite{cit:Moyal} of the Wigner kernel~\cite{cit:Schleich}. The
overall goodness of the expansion is however unclear, since each successive
term in the Moyal expansion contains higher derivatives of both the
distribution and the Hamiltonian.

The modification of the classical mechanics by quantum fluctuations is however
precisely the intuitive understanding associated with Feynman's path integral
description~\cite{cit:Feynman} of quantum mechanics. By adopting a
saddle-point expansion~\cite{cit:Kleinert} one can for example derive the
semiclassical Van Vleck-Gutzwiller propagator~\cite{cit:VanVleck,
cit:Gutzwiller, cit:Heller}. By construction, the quantum fluctuations around
the Euler-Lagrange equations associated with the saddle-point are minimal. In
that respect it is important to note that the Euler-Lagrange equations for the
saddle-point expansion of the Wigner distribution propagator
\cite{cit:SelsBrosensMagnus} differ from the classical equations of motion,
unless the problem is harmonic. Although the classical equation of motion does
lie on a saddle-point, there are infinitely many others such that it has zero
measure in the semiclassical path integral. We therefore expect that there
must exist a modified classical phase space evolution around which the overall
quantum fluctuations are smaller.

In the next section we present a method for deriving an effective classical
Hamiltonian which generates a classical phase space evolution, that is optimal
in a well-defined sense. In section~\ref{sec:quartic}, the method is
illustrated by finding this optimal Hamiltonian for the case of a quartic
oscillator which starts evolving from a Gaussian Wigner function. The result
is shown to compare favorably with the standard truncated Wigner
approximation. Finally, some conclusions are drawn in
section~\ref{sec:Conclusion}.

\section{Optimizing the classical state space evolution}

Consider a $d$-dimensional quantum system, whose initial state at time $0$ is
represented by the density matrix $\hat{\rho}_{0}.$ Under the influence of a
time-dependent Hamiltonian $\hat{H}\left(  t\right)  ,$ the density matrix
$\hat{\rho}(t)$ evolves unitarily as $\hat{\rho}(t)=\hat{U}\left(  t\right)
\hat{\rho}_{0}\hat{U}^{\dagger}\left(  t\right)  $, with
\begin{equation}
\hat{U}\left(  t\right)  =\mathcal{\hat{T}}\exp \left(  -\frac{i}{\hbar}%
{\displaystyle \int \limits_{0}^{t}}
\hat{H}(\tau)\mathrm{d}\tau \right)  ,
\end{equation}
where $\mathcal{\hat{T}}$ denotes the time-ordering operator.

Consider furthermore a set $\boldsymbol{\Sigma}$ of trial density matrices
$\hat{\sigma}(t),$ all generated from the same initial $\hat{\rho}_{0}$ but by
other (approximate) protocols $\Sigma$. It would then be interesting to find
out how each of those protocols $\Sigma$ is performing by comparing
$\hat{\sigma}(t)$ to $\hat{\rho}(t).$ The goodness of a certain protocol
$\Sigma$ can be assessed by measuring the distance $D$ between the two states
underlying $\hat{\sigma}(t)$ and $\hat{\rho}(t),$ a problem that is well
established in the field of quantum information theory~\cite{cit:MWilde}. The
optimal protocol $\Sigma_{\mathrm{opt}}$ is the one that generates the state
closest to the real state, implying%
\begin{equation}
D\left(  \hat{\sigma}_{\mathrm{opt}},\hat{\rho}\right)  =\min_{\Sigma
\in \boldsymbol{\Sigma}}D\left(  \hat{\sigma},\hat{\rho}\right)  .
\end{equation}
However, in quantum information theory the protocols are usually noisy
approximate implementations of the ideal evolution $\hat{U},$ from which one
tries to find the optimal implementation. Here we are concerned with a
different type of approach. In quantum dynamics, the time evolution $\hat
{U}\left(  t\right)  $ is usually analytically and/or numerically
inaccessible. But soluble trial Hamiltonians provide approximation schemes
$\Sigma$ which allow to calculate $\hat{\sigma}(t).$ The distance then
provides a measure for the quality of each approximation. For this purpose we
propose the Hilbert-Schmidt distance measure, i.e.,
\begin{equation}
D_{2}(\hat{\rho},\hat{\sigma})=\sqrt{\mathrm{Tr}\left[  \left(  \hat{\rho
}-\hat{\sigma}\right)  ^{\dagger}\left(  \hat{\rho}-\hat{\sigma}\right)
\right]  }, \label{eq:OptimalH:HilbertSchmittD2}%
\end{equation}
since it is analytically better tractable than the more important trace
distance. So far this is still completely general and one can put every
possibly conceivable approximation in $\boldsymbol{\Sigma}.$ We now
specifically turn to the problem of finding the optimal classical evolution of
the Wigner distribution in phase space.

The Wigner distribution is a joint quasi-probability distribution of canonical
conjugate variables, denoted by $\left(  \mathbf{x,p}\right)  =\left(
\left \{  x_{i}\right \}  ,\left \{  p_{i}\right \}  \right)  $. Although the
Wigner function has the correct marginal probability distributions, it is a
quasi distribution as it can attain negative values. These negative values are
the reason why one must use a quantum measure for the distance between states,
rather than the classical Kullback-Leibler distance between the Wigner
distributions. Denoting the Wigner functions $f_{\rho}(\mathbf{x}%
,\mathbf{p},t)$ and $f_{\sigma}(\mathbf{x},\mathbf{p},t)$ associated with the
states underlying $\hat{\rho}\left(  t\right)  $ and $\hat{\sigma}\left(
t\right)  $ respectively, the Hilbert-Schmidt distance between the two states
becomes%
\begin{equation}
\frac{D_{2}^{2}(\hat{\rho},\hat{\sigma},t)}{2\pi \hbar}=%
{\displaystyle \iint}
\mathrm{d}\mathbf{x}\mathrm{d}\mathbf{p}\left(  f_{\rho}(\mathbf{x}%
,\mathbf{p},t)-f_{\sigma}(\mathbf{x},\mathbf{p},t)\right)  ^{2}.
\label{eq:OptimalH:HilbertSchmittDistance_f}%
\end{equation}
Initially this distance is zero, because then both compared states are equal.
In order to assess the goodness of the classical approximation $f_{\sigma}$
one should investigate the increase of $D_{2}$ over time. The time evolution
of the exact Wigner function $f_{\rho}$ associated with the unitary evolution
$\hat{U}\left(  t\right)  $ is given by%
\begin{equation}
\frac{\partial f_{\rho}\left(  \mathbf{x},\mathbf{p},t\right)  }{\partial
t}-\left \{  H\left(  \mathbf{x},\mathbf{p},t\right)  ,f_{\rho}\left(
\mathbf{x},\mathbf{p},t\right)  \right \}  _{M}=0, \label{eq:OptimalH:WLE}%
\end{equation}
where $\left \{  \cdot,\cdot \right \}  _{M}$ denotes the Moyal
bracket~\cite{cit:Moyal}, and where $H\left(  \mathbf{x},\mathbf{p},t\right)
$ is the Weyl representation~\cite{cit:Hillery, cit:Polkovnikov2010} of the
Hamiltonian $\hat{H}\left(  t\right)  $. The Moyal bracket has the interesting
property that
\begin{equation}
\left \{  \cdot,\cdot \right \}  _{M}=\left \{  \cdot,\cdot \right \}  +O(\hbar
^{2}), \label{sec:OptimalH:Moyal_Poisson}%
\end{equation}
where $\left \{  \cdot,\cdot \right \}  $ is the Poisson bracket. This expansion
forms the basis of the semiclassical truncated Wigner
approximation~\cite{cit:Polkovnikov2010}. By neglecting the quantum
fluctuations of $O(\hbar^{2}),$ the problem of propagating the initial Wigner
distribution in time becomes classical. The corrections around the Poisson
bracket are however higher powers of the Poisson bracket. Therefore, the
overall quality of the approximation will depend on the smoothness of the
distributions on which it acts. It is easy to show that the Moyal bracket and
the Poisson bracket of any quadratic function with any arbitrary distribution
equal each other. As a consequence the time evolution of any state under a
quadratic Hamiltonian is identical to its classical time evolution in phase space.

Instead of approximating the time evolution of the system by its classical
time evolution, consider a classical Hamiltonian $H_{\sigma}$ which differs
from the original one. The corresponding state evolves according to
\begin{equation}
\frac{\partial f_{\sigma}}{\partial t}-\left \{  H_{\sigma},f_{\sigma}\right \}
=0.\label{eq:OptimalH:LE_approximate}%
\end{equation}
Although the full problem~(\ref{eq:OptimalH:WLE}) is supposed not to be
soluble, we can expand $f_{\rho}$ around the approximate solution $f_{\sigma}%
$. In fact, if $f_{\sigma}$ is the optimal approximation of $f_{\rho}$ there
is no better way to estimate the real solution. Up to lowest order, hence in
the short time limit, we find that the distance between the two solutions
grows with time according to
\[
\frac{D_{2}^{2}(\hat{\rho},\hat{\sigma},t)}{2\pi \hbar}=%
{\displaystyle \iint}
\mathrm{d}\mathbf{x}\mathrm{d}\mathbf{p}\left(  \int_{0}^{t}\mathrm{d}%
\tau \left(
\begin{array}
[c]{c}%
\left \{  H,f_{\sigma}\right \}  _{M}\\
-\left \{  H_{\sigma},f_{\sigma}\right \}
\end{array}
\right)  \right)  ^{2}.
\]
The minimization of this distance results in the following requirement for the
effective Hamiltonian $H_{\mathrm{eff}}$ of the optimal classical evolution%
\[
\int_{0}^{t}\mathrm{d}\tau \left \{  \left \{  H_{\mathrm{eff}},f_{\sigma
}\right \}  ,f_{\sigma}\right \}  =\int_{0}^{t}\mathrm{d}\tau \left \{  \left \{
H,f_{\sigma}\right \}  _{M},f_{\sigma}\right \}  .
\]
We would of course like to have the optimal solution for all final times $t.$
Moreover we would like the Hamiltonian to depend locally on time, such that
the approximate evolution (\ref{eq:OptimalH:LE_approximate}) stays Markovian.
Consequently we have to equate the previous expression time by time which
yields the following Euler-Lagrange equation for the optimal Hamiltonian%
\begin{equation}
\left \{  \left \{  H_{\mathrm{eff}},f_{\sigma}\right \}  ,f_{\sigma}\right \}
=\left \{  \left \{  H,f_{\sigma}\right \}  _{M},f_{\sigma}\right \}
.\label{eq:OptimalH:Euler-Lagrange}%
\end{equation}
Since the approximate evolution has to satisfy
Eq.~(\ref{eq:OptimalH:LE_approximate}), this equation can also be written as%
\[
\left \{  \frac{\partial f_{\sigma}}{\partial t}-\left \{  H_{\mathrm{eff}%
},f_{\sigma}\right \}  _{M},f_{\sigma}\right \}  =0.
\]
This implies that the equation of motion~(\ref{eq:OptimalH:LE_approximate})
for $f_{\sigma}$ differs from the equation of motion~(\ref{eq:OptimalH:WLE})
of the true quantum system by an effective quantum jump distribution:%
\begin{align}
\frac{\partial f_{\sigma}}{\partial t}-\left \{  H,f_{\sigma}\right \}  _{M} &
=\left(  \frac{\partial f_{\sigma}}{\partial t}\right)  _{\mathrm{jump}}\\
\text{with }\left(  \frac{\partial f_{\sigma}}{\partial t}\right)
_{\mathrm{jump}} &  =\left \{  H_{\mathrm{eff}},f_{\sigma}\right \}  -\left \{
H,f_{\sigma}\right \}  _{M}.\label{eq:OptimalH:f_jump}%
\end{align}
If $\left \{  \left(  \partial_{t}f_{\sigma}\right)  _{\mathrm{jump}}%
,f_{\sigma}\right \}  =0$ and if $H_{\mathrm{eff}}$ is a solution of
Eq.~(\ref{eq:OptimalH:Euler-Lagrange}), the optimal trial distribution
$f_{\sigma}$ evolves classically. Possible implications of this condition that
the jump distribution and the trial distribution are in mutual involution are
under current investigation.

It is clear that Eq.~(\ref{eq:OptimalH:Euler-Lagrange}) can not be solved in
general. A possible way to simplify the problem is to parametrize the model
Hamiltonian in order to include specific additional physics. In this way the
minimization of the distance will not result in a differential equation for
the Hamiltonian but it will give a set of algebraic equations for the
undetermined parameters in the model Hamiltonian. In \cite{cit:SBMFeff} we
contributed to this possibility in the context of quantum transport. In that
work we minimized the distance (\ref{eq:OptimalH:HilbertSchmittDistance_f})
under the ansatz that the effective Hamiltonian
\[
H_{\mathrm{eff}}(x,p)=\frac{p^{2}}{2m}+V_{\mathrm{eff}}(x),
\]
is parametrized by a scalar potential $V_{\mathrm{eff}}(x).$

Another option is to look for a solution of
Eq.~(\ref{eq:OptimalH:Euler-Lagrange}) for specific simple Wigner functions.
Below we illustrate this approach for the quartic oscillator. Although of
limited practical interest, this example provides the required background for
treating the contact potential in field theory.

\section{Gaussian initial state for a quartic oscillator\label{sec:quartic}}

As an example, consider a one dimensional quartic oscillator%
\begin{equation}
H=\frac{p^{2}}{2}+\frac{x^{2}}{2}+\frac{gx^{4}}{4},\label{eq:quartic:H}%
\end{equation}
expressed in natural units of the harmonic oscillator, i.e., $\hbar
=m=\omega=1.$ In this case the Moyal expansion has a finite number of terms
and becomes%
\begin{align}
\left \{  H,f\right \}  _{M} &  =\left \{  H,f\right \}  -\frac{gx}{4}%
\frac{\partial^{3}f}{\partial p^{3}},\label{eq:quartic:H,f_Moyal}\\
\text{with }\left \{  H,f\right \}   &  =\left(  x+gx^{3}\right)  \frac{\partial
f}{\partial p}-p\frac{\partial f}{\partial x},\label{eq:quartic:H,f_Poisson}%
\end{align}
Since the Moyal bracket is expanded in the Poisson bracket plus a correction,
it is natural to expand the trial Hamiltonian as%
\begin{equation}
H_{\mathrm{eff}}=H+H_{c},\label{eq:quartic:Hc_define}%
\end{equation}
in which $H_{c}$ accounts for the correction and satisfies the following
partial differential equation%
\begin{equation}
\left \{  \left \{  H_{c},f_{\sigma}\right \}  ,f_{\sigma}\right \}  =-\frac{g}%
{4}\left \{  x\frac{\partial^{3}f_{\sigma}}{\partial p^{3}},f_{\sigma}\right \}
.\label{eq:quartic:Hc_motion}%
\end{equation}

As in any problem of quantum dynamics, the initial state has to given, and can
by no means be calculated. Because of the Heisenberg uncertainty principle, a
reasonable example of an initial distribution function is a Gaussian wave
packet in position and momentum. Although many other trial time evolutions can
be imagined, we here illustrate the optimization method for a displaced
version in time of this initial wave packet:%
\begin{multline}
f_{\sigma}(x,p,t)=\frac{1}{2\pi \sigma_{x}\sigma_{p}}e^{-\frac{\left(
x-x_{0}(t)\right)  ^{2}}{2\sigma_{x}^{2}}-\frac{\left(  p-p_{0}(t)\right)
^{2}}{2\sigma_{p}^{2}}}\\
\text{with }2\sigma_{x}\sigma_{p}\geq1.
\end{multline}
The condition $2\sigma_{x}\sigma_{p}\geq1$ accounts for the uncertainty
relation. With this trial distribution function, the equation
(\ref{eq:quartic:Hc_motion}) becomes a inhomogeneous second-order partial
differential equation in $x$ and $p$, with the solution%
\begin{multline*}
H_{c}=\frac{x^{2}g}{48\sigma_{p}^{2}}\left(  18-x\frac{3x-4x_{0}}{\sigma
_{x}^{2}}-6\frac{\left(  p-p_{0}\right)  ^{2}}{\sigma_{p}^{2}}\right) \\
+u\left(  \frac{\left(  x-x_{0}\right)  ^{2}}{2\sigma_{x}^{2}}+\frac{\left(
p-p_{0}\right)  ^{2}}{2\sigma_{p}^{2}}\right)
\end{multline*}
where $u$ is an arbitrary function. The Poisson bracket of $f_{\sigma}$ with
this function $u$ always vanishes. Therefore, one can set $u=0,$ because it
does not contribute to the dynamics. The optimal effective
Hamiltonian~(\ref{eq:quartic:Hc_define}) along these lines thus becomes
\begin{multline}
H_{\mathrm{eff}}=\frac{p^{2}}{2}+\left(  1+\frac{3g}{4\sigma_{p}^{2}}\right)
\frac{x^{2}}{2}+\left(  1-\frac{1}{\left(  2\sigma_{x}\sigma_{p}\right)  ^{2}%
}\right)  \frac{gx^{4}}{4}\nonumber \\
+\frac{g}{\left(  2\sigma_{x}\sigma_{p}\right)  ^{2}}\frac{x_{0}x^{3}}%
{3}-\frac{g}{4\sigma_{p}^{4}}\frac{x^{2}\left(  p-p_{0}\right)  ^{2}}{2}.
\end{multline}
From this effective Hamiltonian one can calculate the effective jump
distribution~(\ref{eq:OptimalH:f_jump}), which turns out to vanish, i.e.,
$\left(  \partial_{t}f_{\sigma}\right)  _{\mathrm{jump}}=0.$ Consequently the
equation of motion (\ref{eq:OptimalH:f_jump}) of the trial $f_{\sigma}$ is
identical to the equation of motion (\ref{eq:OptimalH:WLE}) of the true state
$f_{\rho}$. The non-linearity of the equations of motion will however distort
the Gaussian such that the ansatz state starts to deviate from the real state.

The characteristics of Eq.~(\ref{eq:OptimalH:LE_approximate}) satisfy%
\begin{align*}
\frac{dx}{dt} &  =p-\frac{g}{4}\frac{x^{2}\left(  p-p_{0}\right)  }{\sigma
_{p}^{4}},\\
\frac{dp}{dt} &  =-x+\frac{g}{4}\left(  \frac{x\left(  p-p_{0}\right)  ^{2}%
}{\sigma_{p}^{4}}-\frac{3x}{\sigma_{p}^{2}}+x^{2}\frac{x-x_{0}}{\sigma_{x}%
^{2}\sigma_{p}^{2}}-4x^{3}\right)  .
\end{align*}
Clearly the equations of motion for the average phase space position of the
state become%
\begin{align*}
\frac{dx_{0}}{dt} &  =p_{0},\\
\frac{dp_{0}}{dt} &  =-x_{0}\left(  1+\frac{3}{4}\frac{g}{\sigma_{p}^{2}%
}\right)  -gx_{0}^{3},
\end{align*}
which for $\sigma_{p}\rightarrow \infty$ (i.e., $\sigma_{x}\rightarrow0$)
reduce to the equations of motion of the real Hamiltonian~(\ref{eq:quartic:H}%
). The latter equations, including the $\sigma_{p}$-dependent term, express
Ehrenfest's theorem~\cite{citEhrenfest}. 

The restoration of Ehrenfest's theorem is indicative of the gain made by
propagating the classical trial system with the effective Hamiltonian rather
than the real Hamiltonian. It should moreover be noted that, at least for the
present example, the initial behavior of the distance is drastically different
in both situations. Within the truncated Wigner approximation the initial
distance will always increase linear in time. The slope is simply determined
by the initial value of the $O(\hbar^{2})$ correction to the Poisson bracket
in order to obtain the Moyal bracket. So although the correction is of
$O(\hbar^{2}),$ it does cause a discrepancy between the exact and the
truncated distribution which increases linearly in time. The variational
result however has vanishing corrections within first order, which implies
that the initial increase in the distance is quadratic. The result is thus
correct up to second order, i.e., up to the deviation from Gaussianity due to
the non-linearity of the equations of motion. The curvature can not be
estimated within the present approach but can be found by doing perturbation
theory in terms of quantum jumps~\cite{cit:Polkovnikov2010} around the
variational state.

Finally note that the present effective Hamiltonian can be used as an ansatz
effective Hamiltonian for the optimization of states that are not Gaussian.
The efficiency of course greatly depends on the shape of the state but if
properly parametrized the initial time evolution can never be worse then the
truncated Wigner result.

\section{Conclusion\label{sec:Conclusion}}

In conclusion we have presented a method to extract the optimal Hamiltonian
for the classical time evolution of the Wigner distribution associated with
the unitary evolution of an arbitrary initial state. In general this effective
classical Hamiltonian is different from the Hamiltonian which generates the
quantum dynamics. The effective classical Hamiltonian was explicitly
calculated for an initial Gaussian trial state in a quartic oscillator. Unlike
in the truncated Wigner approximation, the equation of motion was shown to be
in agreement with Ehrenfest's theorem and the quantum corrections around the
result are of second order.

%

\end{document}